\documentclass{PoS}

\title{Identified particle production in inelastic $pp$ events with the ATLAS detector}

\ShortTitle{Identified particle production in inelastic $pp$ events with the ATLAS detector}

\author{\speaker{Leonid Gladilin}\thanks{On behalf of the ATLAS Collaboration.}\\
        Moscow State University - Scobeltsyn Institute of Nuclear Physics\\
1(2), Leninskie gory, GSP-1, Moscow 119991 - Russia\\
        E-mail: \email{Leonid.Gladilin@cern.ch}}

\abstract{Various strange and charmed hadrons were reconstructed
with the ATLAS detector in $pp$ collisions at $\sqrt{s}=7\,$TeV.
The data sample was collected in March-May of 2010 using
a minimum-bias trigger.
The $K^0_S$ and $\Lambda^0$ kinematic distributions were
studied using data corresponding to
an integrated luminosity of $190\,\mu$b$^{-1}$.
The $\Xi^\mp$ and $\Omega^\mp$ baryons
were reconstructed in their cascade decays
in data corresponding to
an integrated luminosity of $250\,\mu$b$^{-1}$.
The $D^{*\pm}$, $D^\pm$ and $D_s^\pm$
charmed mesons
were reconstructed in
the range of transverse momentum $p_{\mathrm T}(D^{(*)})>3.5\,$GeV
and pseudorapidity $|\eta(D^{(*)})|<2.1$
in data corresponding to
an integrated luminosity of $1.4\,$nb$^{-1}$.
The fitted mass values were found to be in agreement
with their world averages
while the observed invariant mass resolutions agree with
Monte Carlo expectations.
This study confirms the high performance of the ATLAS detector
for precision tracking measurements.}

\FullConference{35th International Conference of High Energy Physics - ICHEP2010,\\
		July 22-28, 2010\\
		Paris France}

\begin{document}

\section{Introduction}

The reconstruction of identified particles
containing strange and charm quarks is already feasible
with first LHC data
due to the large expected cross sections and clean particle signatures.  
A study of $\phi(1020)\rightarrow K^+K^-$ decays
with the ATLAS detector~\cite{atlas}
in $\sqrt{s}=900\,$GeV collision data has been reported
earlier~\cite{phi,performance}.
The $dE/dx$ particle identification
was used to select charged kaons in that analysis.
Results on identified strange and charmed particles production
in $pp$ collisions at $\sqrt{s}=7\,$TeV are presented in this note.

The data used in this analysis were collected between March and May, 2010.
Events selected with a minimum-bias trigger and with a reconstructed
primary vertex were kept for particle reconstruction.
The $dE/dx$ particle identification
was not used.
Non-diffractive minimum-bias Monte Carlo (MC) was used to tune
the selection criteria and to make comparisons with data.

\section{Strange Particle Production}

\begin{figure}[b]
\includegraphics[width=.5\textwidth]{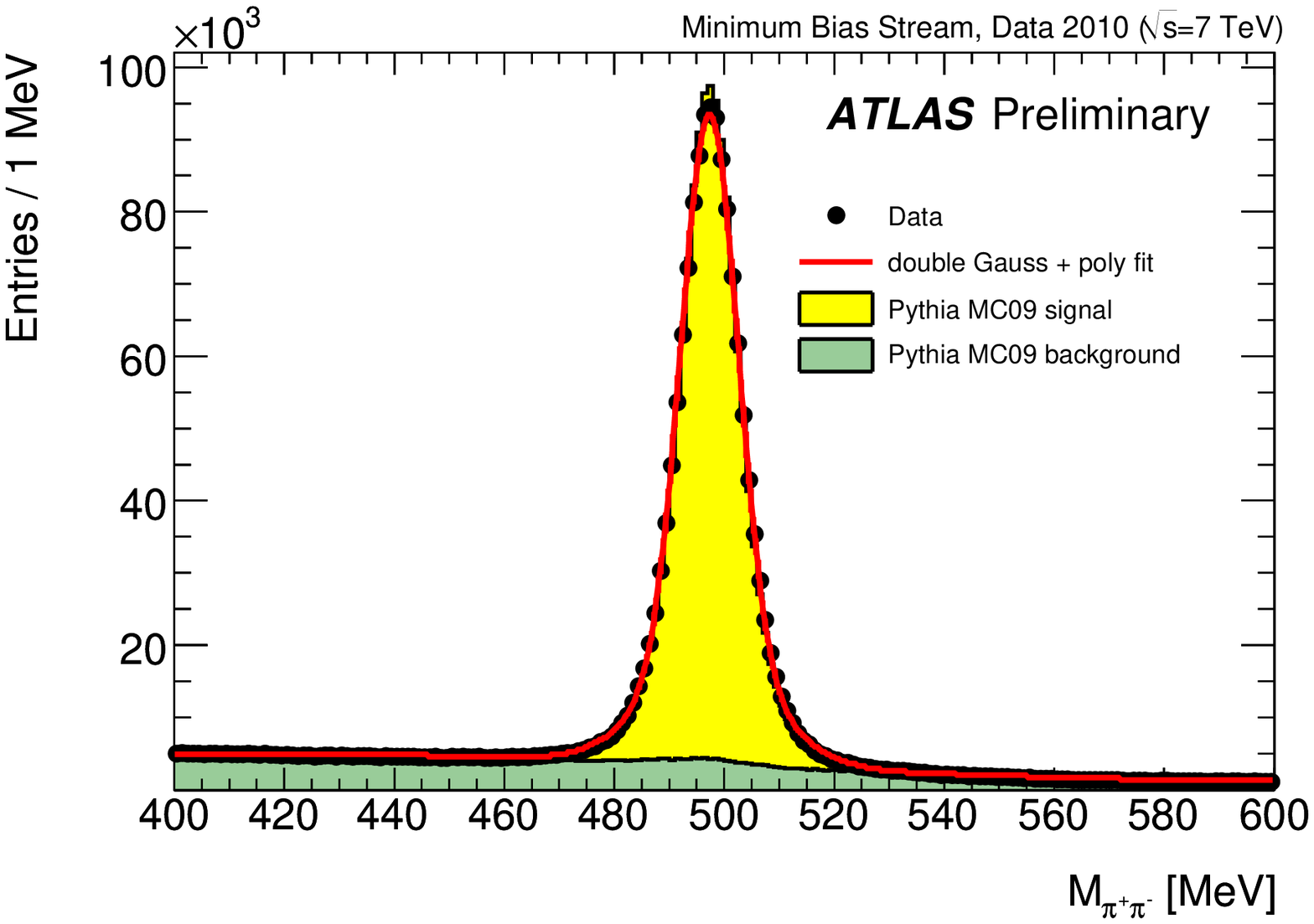}
\includegraphics[width=.5\textwidth]{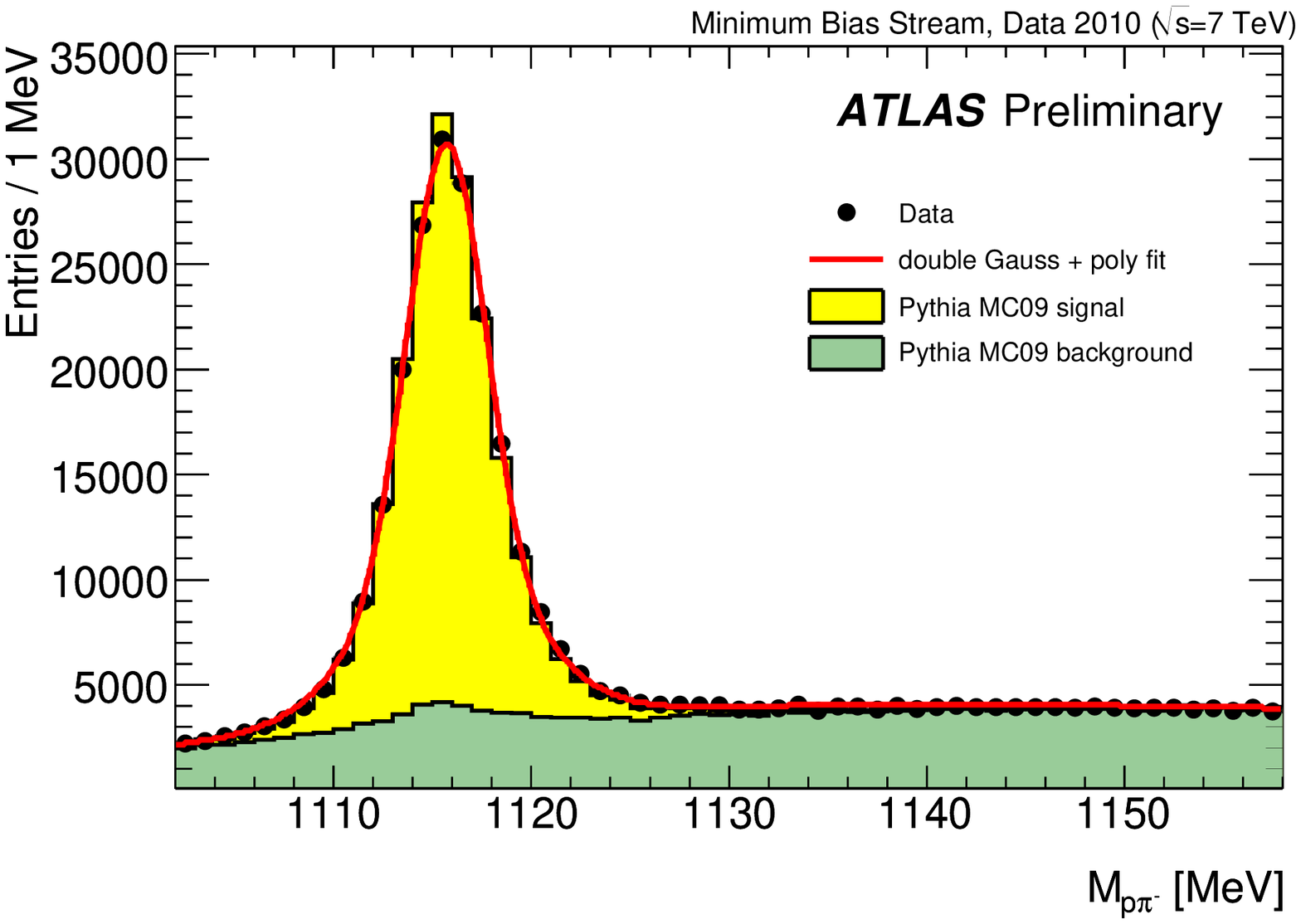}
\caption{The $M(\pi^+\pi^-)$ distribution for $K^0_S$ candidates (left plot)
and $M(p\pi^-)$ for $\Lambda^0$ candidates (right~plot).
The points are data, while the histograms show the MC simulation.
The solid curves represent fit results.
The fitted masses (widths) are
$497.427\pm 0.006\,$MeV ($5.60\,$MeV) and $1115.73\pm 0.01\,$MeV ($2.28\,$MeV)
for the $K^0_S$ and $\Lambda^0$ signals, respectively.
The fit uncertainties on the mass values are statistical only.
}
\label{fig:k0lambda}
\end{figure}

The decays $K^0_S\rightarrow \pi^+\pi^-$, $\Lambda^0\rightarrow p\pi^-$
and ${\bar\Lambda}^0\rightarrow \bar{p}\pi^+$
were reconstructed by fitting pairs of tracks
to a secondary vertex in data corresponding to
an integrated luminosity of $190\,\mu$b$^{-1}$~\cite{k0lambda}.
Only tracks with transverse momentum $p_{\mathrm T}>100\,$MeV were used.
The transverse distance between the secondary and primary vertices
was required to be at least $4\,$mm for $K^0_S$ candidates, while
the minimal there-dimensional distance of $30\,$mm was required
for $\Lambda^0$ and ${\bar\Lambda}^0$ candidates.
The distributions of the invariant mass of the $K^0_S$ and $\Lambda^0$
candidates in data and MC simulation are shown in Fig.~\ref{fig:k0lambda}.
The signal positions and widths are consistent with the MC expectations
and with the world average mass values~\cite{pdg2008}.
Distributions of pseudorapidity and azimuthal angle
for the reconstructed $K^0_S$, $\Lambda^0$ and ${\bar\Lambda}^0$ candidates 
are described reasonably well by the MC simulation,
while the MC has a greater fraction of candidates
at higher values of the candidate transverse momentum~\cite{k0lambda}.

\begin{figure}
\includegraphics[width=.5\textwidth]{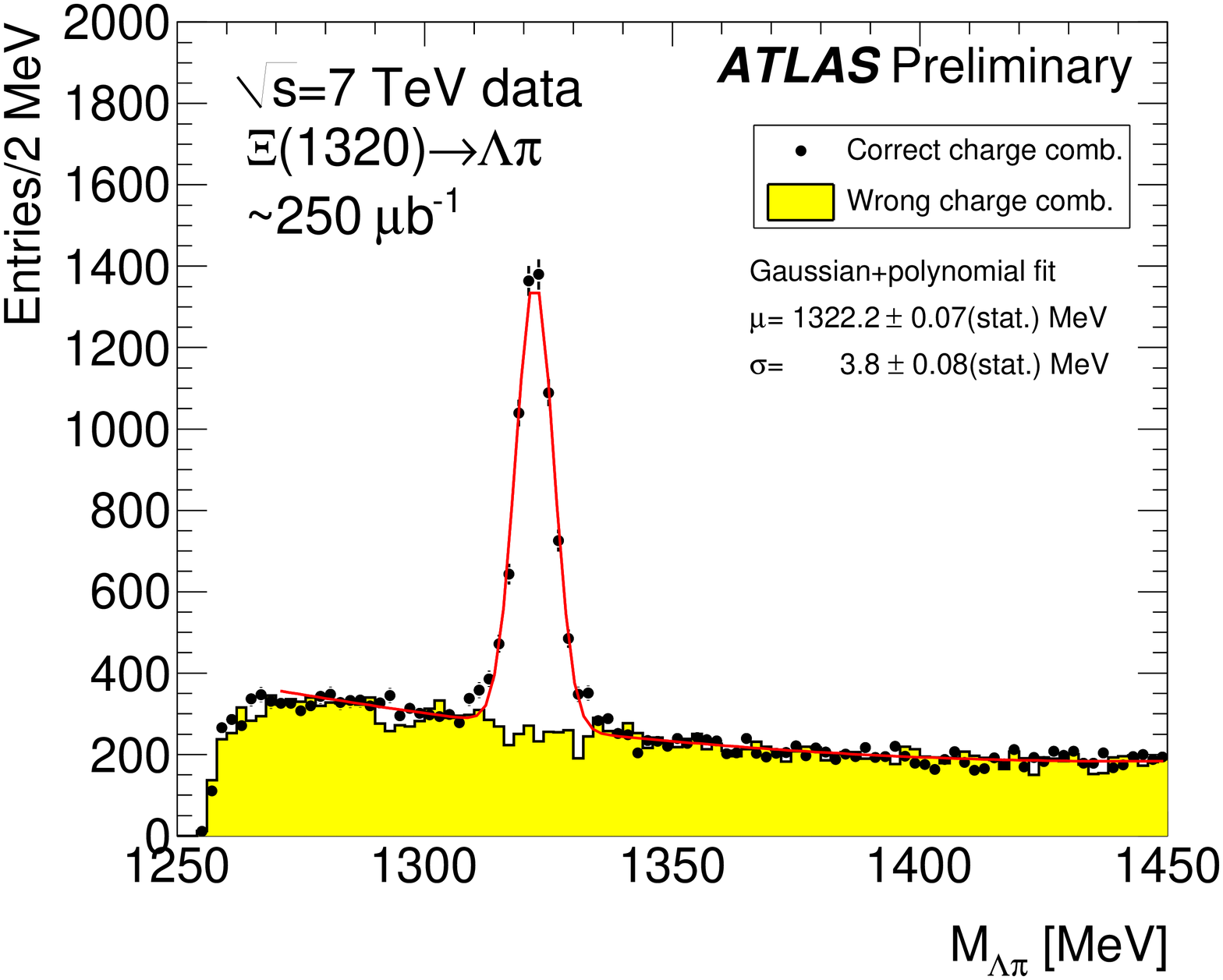}
\includegraphics[width=.5\textwidth]{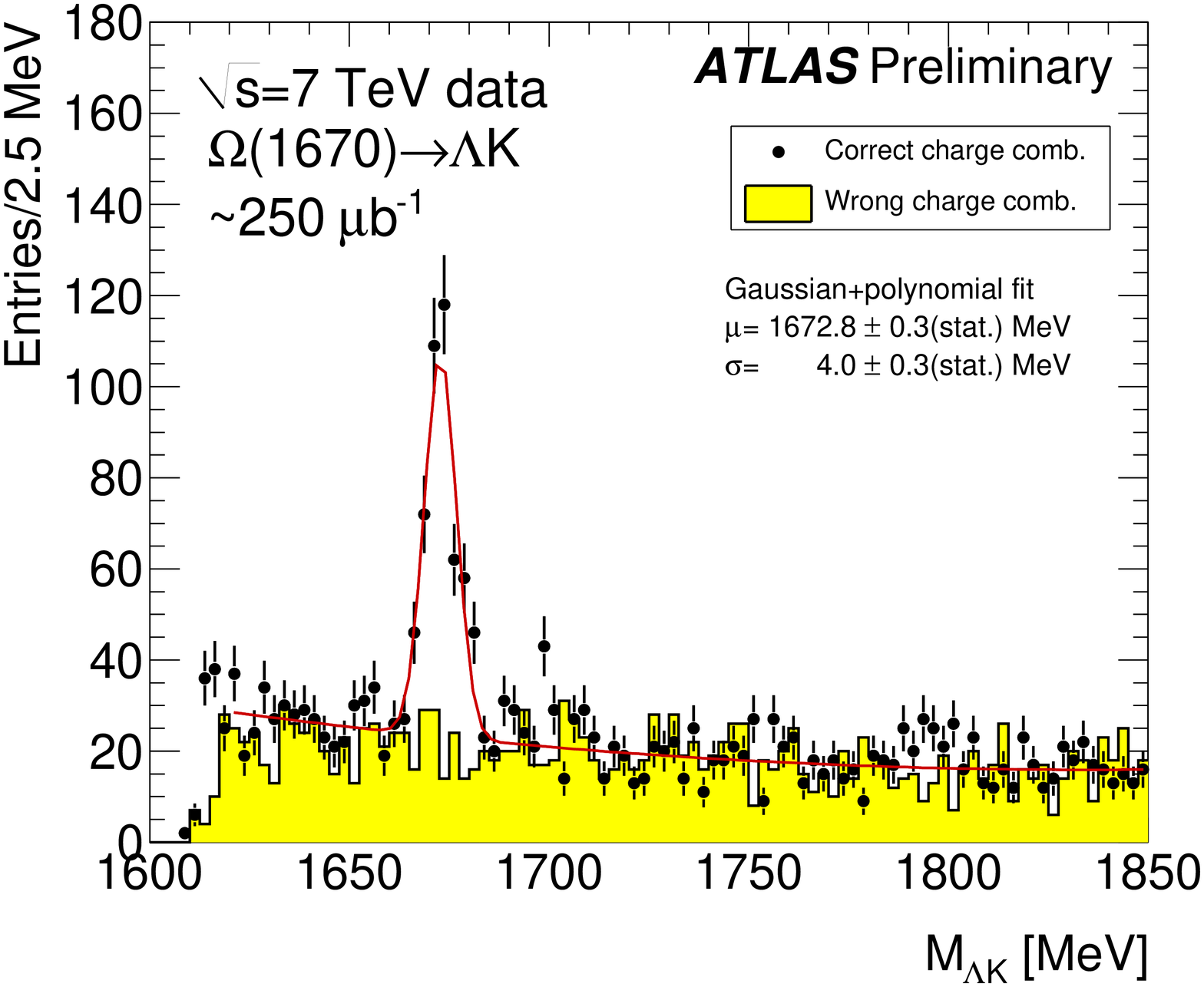}
\caption{The $M(\Lambda\pi)$ distribution for $\Xi^\mp$ candidates (left plot)
and $M(\Lambda K)$ for $\Omega^\mp$ candidates (right plot).
The histograms show the distributions for wrong-charge combinations.
The solid curves represent fit results.}
\label{fig:cascades}
\end{figure}

The cascade decays $\Xi^-\rightarrow \Lambda^0\pi^-$
and $\Omega^-\rightarrow \Lambda^0 K^-$ with
$\Lambda^0\rightarrow p\pi^-$ (+ c.c.)
were reconstructed by fitting
secondary and tertiary vertices
in data corresponding to
an integrated luminosity of $250\,\mu$b$^{-1}$~\cite{cascades}.
The additional track for the first cascade vertex
was required to have transverse momentum above $150\,$MeV
for $\Xi^\mp$ candidates and above $400\,$MeV
for $\Omega^\mp$ candidates.
The transverse distance between the first cascade vertex and primary vertex
was required to be at least $4\,$mm for the $\Xi^\mp$ candidates
and at least $6\,$mm for the $\Omega^\mp$ candidates.
The distributions of the invariant mass of the $\Xi^\mp$ and $\Omega^\mp$
candidates in data and MC simulation are shown in Fig.~\ref{fig:cascades}.
The signal positions and widths are consistent with the MC expectations
and with the world average mass values~\cite{pdg2008}.

\section{Charmed Particle Production}

The $D^{*\pm}$, $D^\pm$ and $D_s^\pm$
charmed mesons
were reconstructed in
the range of transverse momentum $p_{\mathrm T}(D^{(*)})>3.5\,$GeV
and pseudorapidity $|\eta(D^{(*)})|<2.1$
in data corresponding to
an integrated luminosity of $1.4\,$nb$^{-1}$~\cite{dmesons}.

\begin{figure}
\includegraphics[width=.5\textwidth]{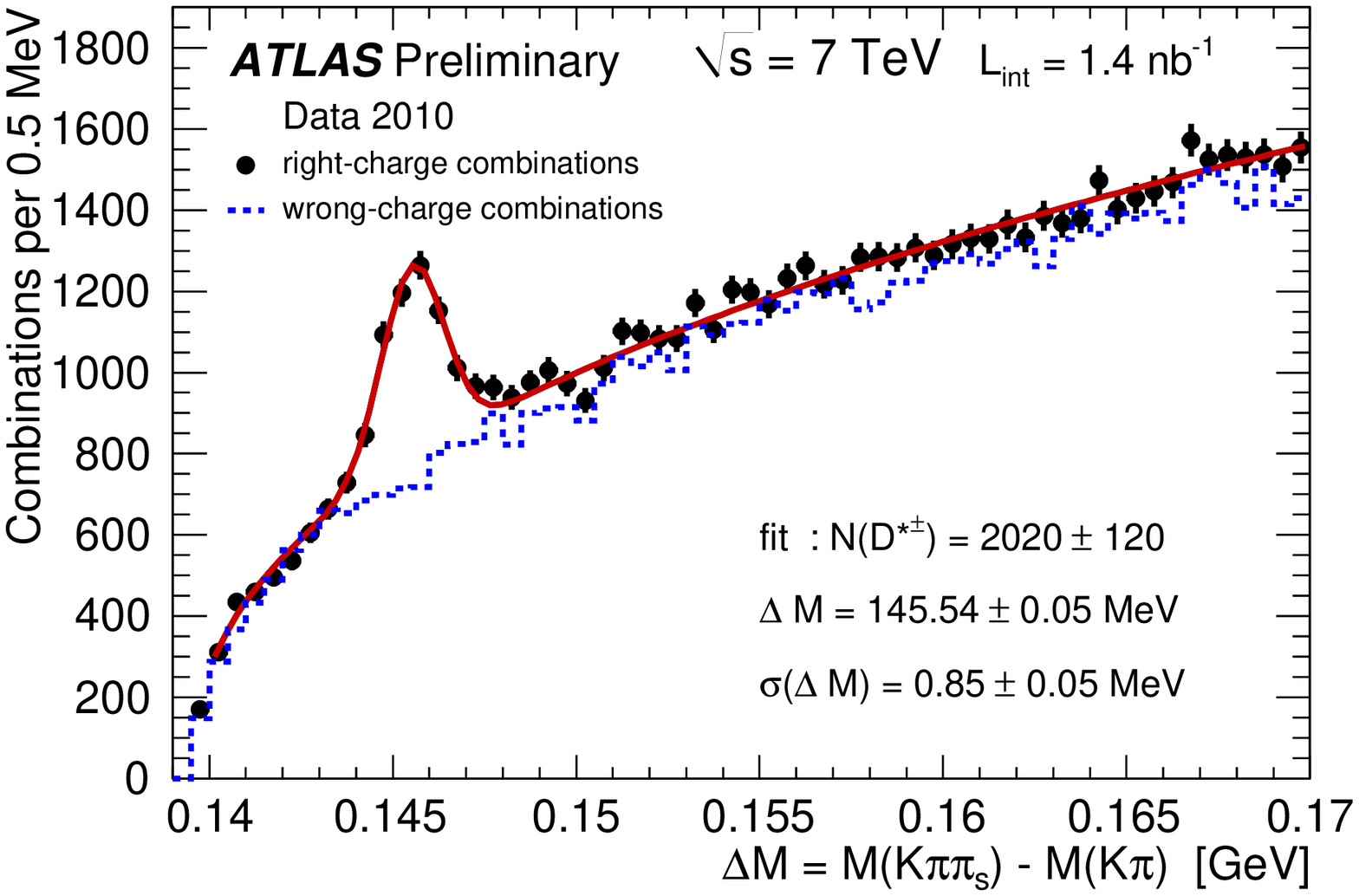}
\includegraphics[width=.5\textwidth]{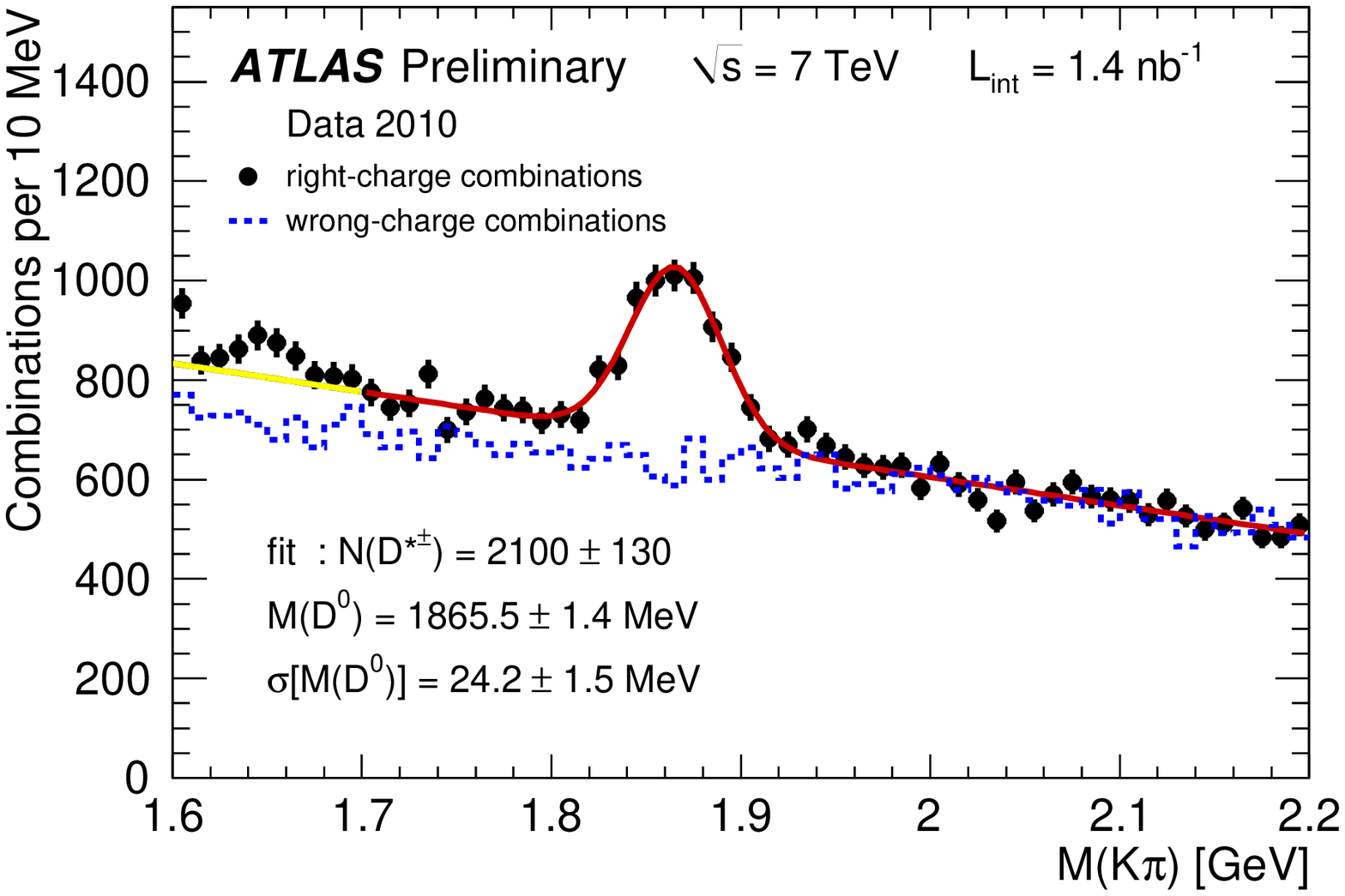}
\vspace*{-0.6cm}
\caption{
The distribution of the mass difference,
$\Delta M=M(K \pi \pi_s)-M(K \pi)$, (left plot)
and the $M(K \pi)$ distribution (right plot)
for the $D^{*\pm}$  candidates (points).
The dashed histograms
show the distributions for wrong-charge combinations.
The solid curves represent fit results.}
\label{fig:dstar}
\end{figure}

The $D^{*\pm}$ mesons were identified using the decay channel
$D^+\rightarrow D^0\pi^{+}_{s}\rightarrow(K^{-}\pi^{+})\pi^{+}_{s}$,
where the pion from the $D^{*+}$ decay is referred to as
the ``soft'' pion, $\pi_s$.
The transverse decay length of the $D^0$ candidates was required
to satisfy $L_{\rm XY}>0$.
The left plot in Fig.~\ref{fig:dstar} shows
the mass difference, $\Delta M=M(K \pi \pi_s)-M(K \pi)$,
distribution for the
$D^{*\pm}$ candidates
which satisfy $1.83<M(K \pi)<1.90\,$GeV,
while the right plot in Fig.~\ref{fig:dstar} shows
the $M(K\pi)$ distribution
for the
$D^{*\pm}$ candidates
which satisfy $144<\Delta M<147\,$MeV.
The fitted $D^{*\pm}$ yield was
$N(D^{*\pm})=2020\pm120$.

\begin{figure}
\includegraphics[width=.5\textwidth]{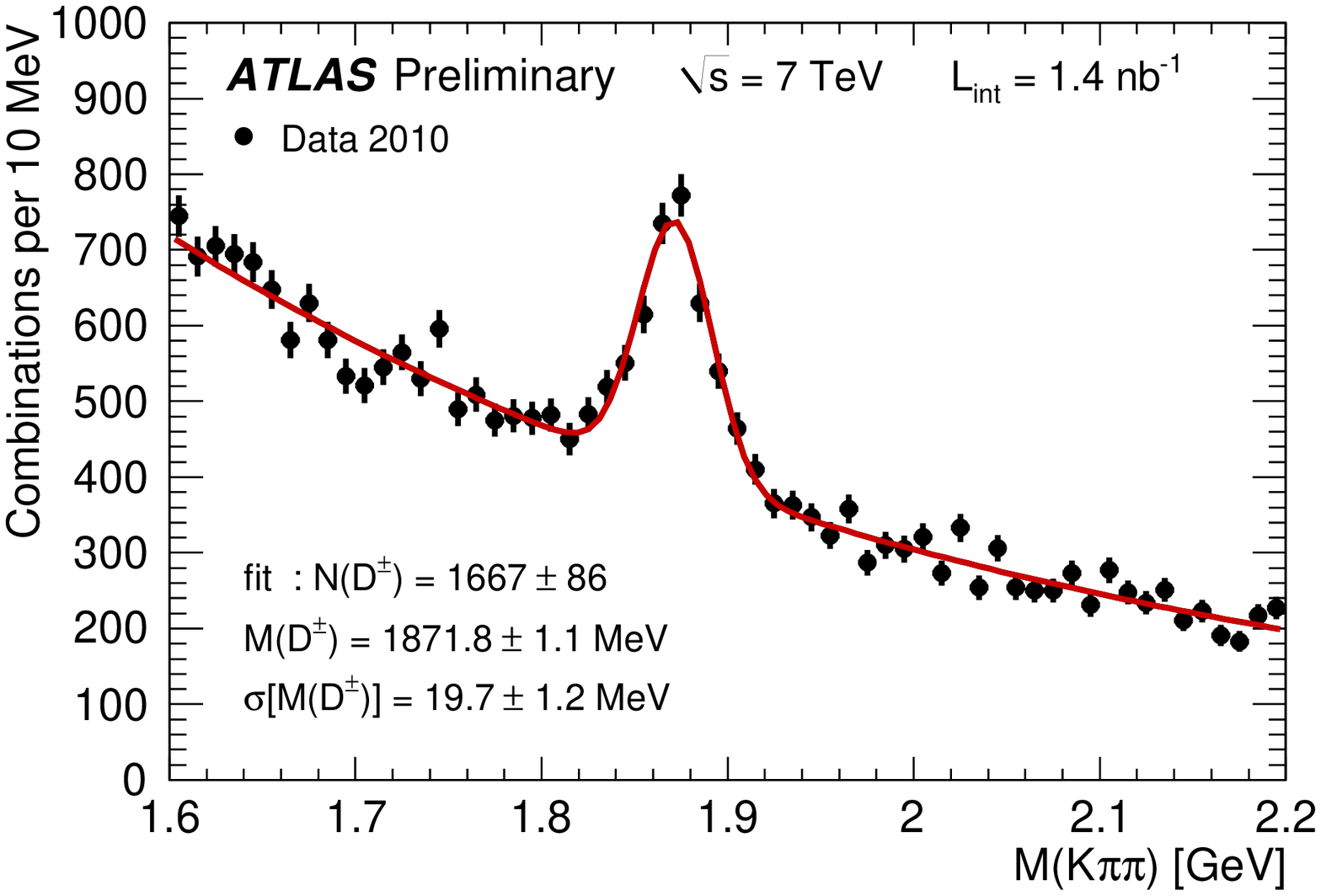}
\includegraphics[width=.5\textwidth]{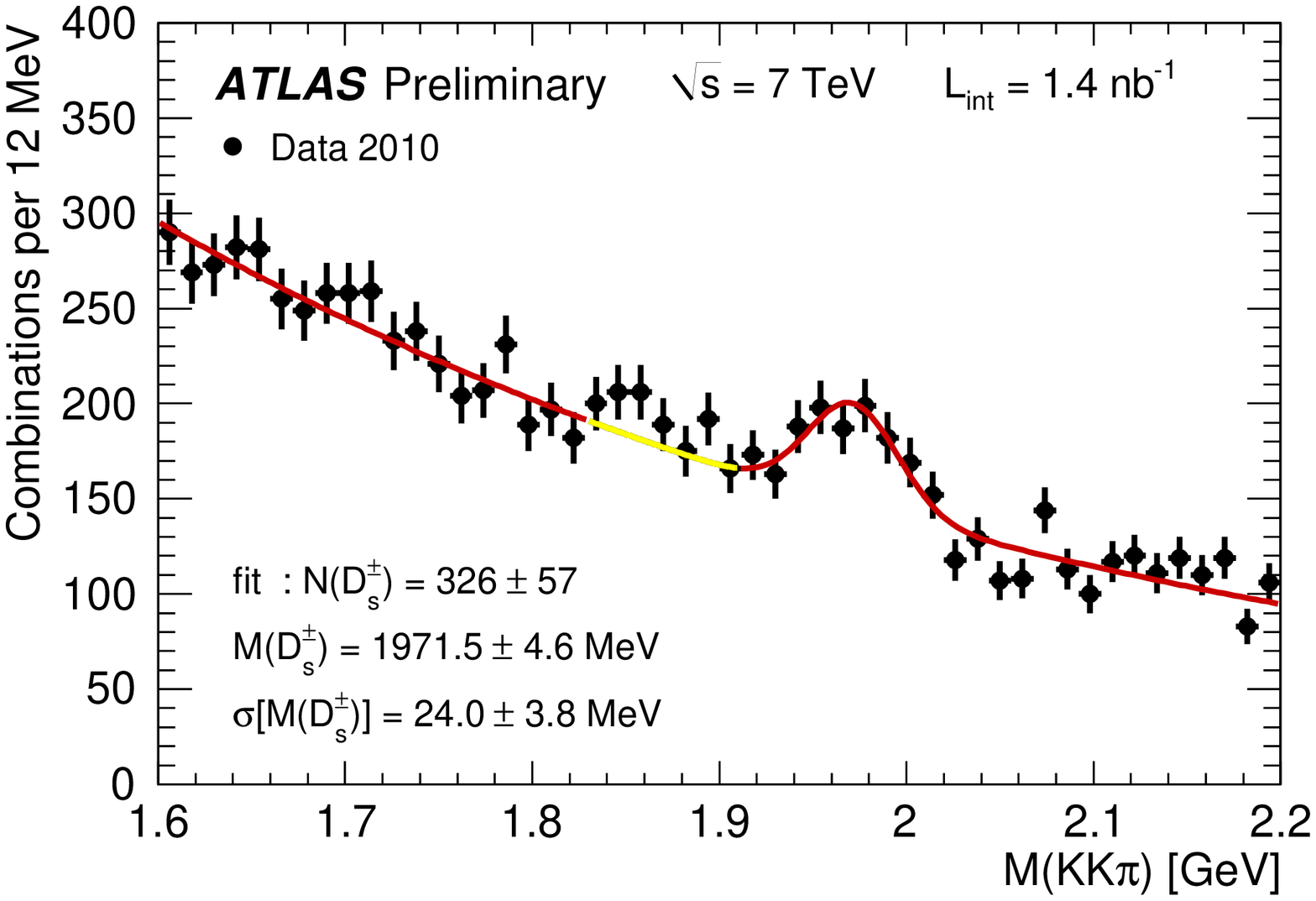}
\vspace*{-0.6cm}
\caption{
The $M(K \pi \pi)$ distribution
for the $D^{\pm}$  candidates (left plot)
and the $M(K K \pi)$ distribution
for the $D_s^{\pm}$  candidates (right plot).
The solid curves represent fit results.}
\label{fig:dcdss}
\end{figure}

The $D^\pm$ mesons were reconstructed
from the decay
$D^+ \rightarrow K^-\pi^+\pi^+$,
and $D_s^\pm$ mesons were reconstructed
from the decay
$D_s^+ \rightarrow \phi\pi^+$ with $\phi \rightarrow K^+K^-$.
The transverse decay lengths of the $D^\pm$ and $D_s^\pm$ candidates
were required
to be above $1.3\,$mm and $0.4\,$mm, respectively.
For $D_s^\pm$ candidates, the $M(KK)$ invariant mass was required to be
within $\pm6\,$MeV of the $\phi$ mass.
The $M(K \pi \pi)$ distribution
for the $D^{\pm}$  candidates
and the $M(K K \pi)$ distribution
for the $D_s^{\pm}$  candidates
are shown in Fig.~\ref{fig:dcdss}.
The fitted yields were
$N(D^\pm)=1667\pm86$ and $N(D_s^\pm)=326\pm57$.

The fitted masses of the reconstructed charmed mesons
were found to be in agreement
with their world averages~\cite{pdg2008}
while the observed invariant mass resolutions agree with MC
expectations.


\end{document}